\documentclass[runningheads]{llncs}
\usepackage{graphicx}
\usepackage{amsmath}
\usepackage{xcolor}
\usepackage{graphicx}
\usepackage{wrapfig}
\usepackage{psfrag}
\usepackage{afterpage}
\usepackage{subfigure}
\usepackage{comment}
\usepackage{pifont}
\newcommand{\cmark}{\ding{51}}%
\newcommand{\xmark}{\ding{55}}%

\begin{document}
\title{Putting Data Science Pipelines on the Edge
}
%
%
\author{Ali Akoglu\inst{1} \and
Genoveva Vargas-Solar\inst{2}\thanks{Authors list is given in alphabetical order.}} 

\authorrunning{A. Akoglu, G. Vargas-Solar}
%
\institute{ECE, University of Arizona \\ Tucson, AZ, USA\\
\email{akoglu@arizona.edu}
\and
French Council of Scientific Research (CNRS), LIRIS-LAFMIA \\ 69622 Villeurbanne, France \\
\email{genoveva.vargas-solar@liris.cnrs.fr} 
}
\maketitle              
\begin{abstract}
This paper proposes a composable "Just in Time Architecture" for Data Science (DS) Pipelines named JITA-4DS  and associated resource management techniques for configuring disaggregated data centers (DCs).  DCs under our approach are composable based on vertical integration of the application, middleware/operating system, and hardware layers
customized dynamically to meet application Service Level Objectives  (SLO - application-aware management). Thereby, pipelines utilize a set of flexible building blocks that can be dynamically and automatically assembled and re-assembled to meet the dynamic changes in the workload's SLOs. To assess disaggregated DC's, we study how to model and validate their performance in large-scale settings.
\vspace{-4mm}
\end{abstract}
%
%
\section{Introduction}
Data infrastructures such as Google, Amazon, eBay, and E-Trade are powered by data centers (DCs) with tens to hundreds of thousands of computers and storage devices running complex software applications. 
Since 2018, the word-wide spending in big data solutions and public cloud infrastructure has increased by 12\% and 24\%, respectively~\cite{idcapril19}.
Existing IT architectures are not designed to provide an agile infrastructure to keep up with the rapidly evolving next-generation mobile, big data, and data science pipelines demands. These applications are distinct from the "traditional" enterprise ones because of their size, dynamic behavior, and nonlinear scaling and relatively unpredictable growth as inputs being processed. Thus, they require continuous provisioning and re-provisioning of DC resources~\cite{chen19acm,kannan19eurosys,xu2015enreal} given their dynamic and unpredictable changes in the Service Level Objectives (SLOs) (e.g., availability response time, reliability, energy).   
Computer manufacturers, software developers, and service providers cannot cope with the dynamic and continuous changes in applications and workload types. Consequently, their offered services become unstable, fragile, and cannot guarantee the required SLOs for other application classes. 

\vspace{-1mm}
This paper targets the execution of data science (DS) pipelines \footnote{A Data Science Pipeline consists of a set of data processing tasks organised as a data flow defining the data dependencies among the tasks and a control flow defining the order in which tasks are executed.} supported by data processing, transmission and sharing across several resources executing greedy processes. Current data science pipelines environments promote high-performance cloud platforms as backend support for completely externalising their execution. These platforms provide various infrastructure services with compute resources such as general purpose processors (GPP), Graphics Processing Units (GPUs), Field Programmable Gate Arrays (FPGAs) and Tensor Processing Unit (TPU) coupled with platform and software services to design, run and maintain DS pipelines. These one-fits-all solutions impose the complete externalisation of data pipeline tasks that assume (i) reliable and completely available network connection; (ii) can be energy and economically consuming, allocating large scale resources for executing pipelines tasks. However, some tasks can be executed in the edge, and the backend can provide just in time resources to ensure ad-hoc and elastic execution environments.

Our research investigates architectural support, system performance metrics, resource management algorithms, and modeling techniques to enable the design of composable (disaggregated) DCs. 
The goal is to design an innovative composable “Just in Time Architecture” for configuring DCs for Data Science Pipelines (JITA-4DS) and associated resource management techniques. DCs  utilize a set of flexible building blocks that can be dynamically and automatically assembled and re-assembled to meet the dynamic changes in workload’s Service Level Objectives (SLO) of current and future DC applications.
DCs under our approach are composable based on vertical integration of the application, middleware/operating system, and hardware layers 
customized dynamically to meet application SLO (application-aware management). 
Thus,  DCs configured using JITA-4DS provide ad-hoc environments efficiently and effectively meeting the continuous changes in requirements of data-driven applications or workloads (e.g., data science pipelines). 
To assess disaggregated DC's, we study how to model and validate their performance in large-scale settings. We rely on novel model-driven resource management heuristics based on metrics that measure a service's value for achieving a balance between competing goals (e.g., completion time and energy consumption). 
Initially, we propose a hierarchical modeling approach that integrates simulation tools and models. 


The remainder of the paper is organised as follows. Section \ref{sec:related} discusses related work identifying certain drawbacks and  issues we believe still remain open. Section \ref{sec:edgepipeline} JITA-4DS the just in time edge based data science pipeline execution environment proposed in this paper. Section \ref{sec:preliminary} describes preliminary results regarding  JITA-4DS. Finally Section \ref{sec:conclusion} concludes the paper and discusses future work.

\section{Related Work}\label{sec:related}

The work introduced in this paper is related to two types of approaches: (i) disaggregated data centers willing to propose alternatives to one fits all architectures; and (ii) data science pipelines' execution platforms relying on cloud services for running greedy data analytics tasks.
\vspace{-3mm}
\paragraph{\bf Disaggregated data centers}
Disaggregation of IT resources has been proposed as an alternative configuration for data centers. Compared to the monolithic server approach, in a disaggregated data center, CPU, memory and storage are separate resource blades interconnected via a network. The critical enabler for the disaggregated data center is the network and management software to create the logical connection of the resources needed by an application \cite{7842314}.
%
The industry has started to introduce systems that support a limited disaggregation capability. For example, the Synergy system by Hewlett Packard Enterprise (HPE) ~\cite{synergy}, and the Unified Computing System (UCS)~\cite{cisco} M series servers by Cisco are two commercial examples of composable infrastructures. 
%
\cite{7842314} proposes a disaggregated data
center network architecture, with a scheduling algorithm designed for disaggregated computing.
\vspace{-3mm}
\paragraph{\bf Data Science Environments}

{\em Data analytics stacks}
environments provide the underlying infrastructure for managing data, implementing data processing workflows to transform them, and executing data analytics operations (statistics, data mining, knowledge discovery, computational science processes). 
For example, the Berkeley Data Analytics Stack (BDAS) from the AMPLAb project is a multi-layered architecture that provides tools for virtualizing resources, addressing storage, data processing and querying as underlying tools for big data-aware applications. AsterixDB from the Asterix project is a full-fledged big data stack designed as a scalable, open-source Big Data Management System (BDMS \url{https://asterixdb.apache.org}).  
%
%

{\em Cloud based Data Science Environments}
provide tools to explore, engineer and analyse data collections. 
They are notebook oriented environments \footnote{A notebook is a JSON document, following a versioned schema, and containing an ordered list of input/output cells which can contain code, text (using Markdown 
mathematics, plots and rich media.} externalised on the cloud. They provide data labs  and  environments with  libraries for defining and executing notebooks.  Examples of existing data labs are Kaggle 
and CoLab 
from Google, and Azure Notebooks from Microsoft Azure. 
%
%
%
{\em Platforms for custom modelling}
provide a suite of machine learning tools allowing developers with little experience  to train high quality models.  Tools are provided as services by commercial cloud providers that include storage, computing support and   environments for training and enacting greedy artificial intelligence (AI) models. The main vendors providing this kind of platforms are Amazon Sage Maker, 
Azure ML Services, 
Google ML Engine 
and IBM Watson ML Studio. 
{\em Machine Learning and Artificial Intelligence Studios}
  give  an interactive, visual workspace to  build, test, and iterate on analytics models and develop experiments \footnote{ 
An experiment consists of data sets that provide data to analytical modules, connected together to construct an analysis model.}.
 An experiment has at least associated one data set and one module.
Data sets may be connected only to modules, and modules may be connected to either data sets or other modules.
All input ports for modules must have some connection to the data flow.
All required parameters for each module must be set.
%
{\em Machine learning runtime environments} provide the tools needed for executing machine learning workflows, including data stores, interpreters and runtime services like Spark, Tensorflow and Caffe for executing analytics operations and models.
The most prominent studios are, for example, Amazon Machine Learning, Microsoft Artificial Intelligence and Machine Learning Studio, Cloud Auto ML,  Data Bricks  ML Flow and IBM Watson ML Builder. 

\vspace{-3mm}
\paragraph{\bf Discussion}
Data Analytics Stacks remain general solutions provided as "one-fits-all" systems with which, of course, big data can be managed and queried through built-in or user-defined operations integrated into imperative or SQL like solutions.
In contrast, data labs' objective is to provide tools for managing and curating data collections, automatically generating qualitative and quantitative meta-data. 
Curated data collections associated with a search engine can be shared and used in target data science projects. Data labs offer storage space often provided by a cloud vendor (e.g., users of CoLab use their google drive storage space for data collections, notebooks and results). Execution environments associate computing resources for executing notebooks that use curated data collections.
Machine learning studios address the analytics and data management divide with integrated backends for dealing with efficient execution of analytics activities pipelines allocating the necessary infrastructure  (CPU, FPGA, GPU, TPU) and platform (Spark, Tensorflow)  services.

These environments provide resources (CPU, storage and main memory) for executing data science tasks. These tasks are repetitive, process different amounts of data and require storage and computing support. Data science projects have life cycle phases that imply in-house small scale execution environments, and they can evolve into deployment phases where they can touch the cloud and the edge resources. Therefore, they require underlying elastic architectures that can provide resources at different scales. Disaggregated data centers solutions seem promising for them. Our work addresses the challenges implied when coupling disaggregated solutions with data science projects. 

\section{JITA-4DS: Just in time Edge Based Data Science Pipelines Execution}\label{sec:edgepipeline}

\begin{figure*}[h]
   \centering
   \includegraphics[width=\linewidth]{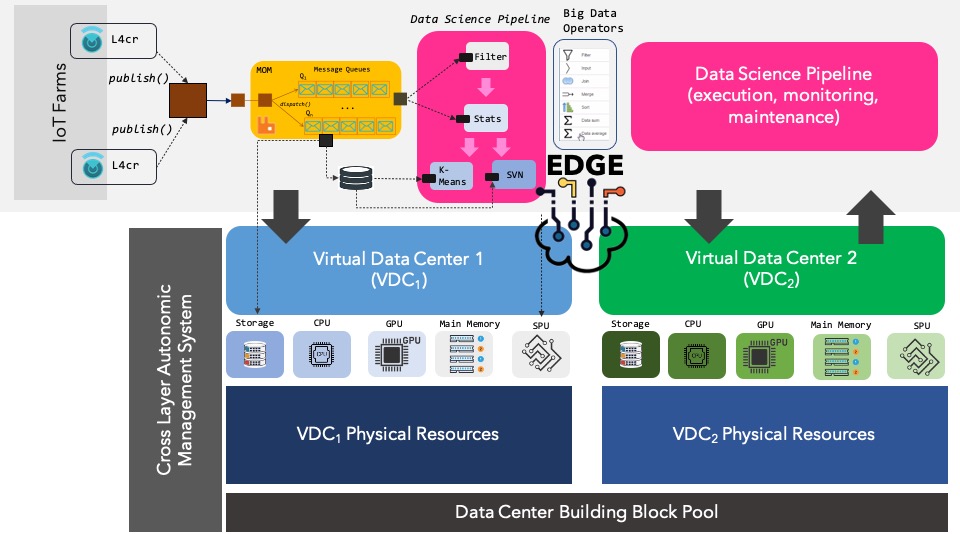}
    \caption{Just in Time Architecture for Data Science Pipelines - JITA-4DS}
   \label{fig:jita}
   \vspace{-4mm}
 \end{figure*}

The Just in Time Architecture for Data Science Pipelines (JITA-4DS), illustrated in Figure \ref{fig:jita}, is a cross-layer management system that is aware of both the application characteristics and the underlying infrastructures to break the barriers between applications, middleware/operating system, and hardware layers. Vertical integration of these layers is needed for building a customizable Virtual Data Center (VDC) to meet the dynamically changing data science pipelines' requirements such as performance, availability, and energy consumption. 

JITA-4DS fully exploits the virtualization from the virtual machine (VM) level into the VDC level  (e.g.,  fine-grain resource monitoring and control capabilities). For the execution of data science pipelines,  JITA-4DS  can build a VDC that can meet the application SLO, such as execution performance and energy consumption. The selected VDC, then, is mapped to a set of heterogeneous computing nodes such as GPPs, GPUs, TPUs, special-purpose units (SPUs) such as ASICs and FPGAs, along with memory and storage units.


DS pipelines running on top of JITA-4DS VDC's apply sets of big data processing operators to stored data and streams produced by the Internet of Things (IoT) farms (see the upper part of Figure \ref{fig:jita}). In the JITA-4DS approach, the tasks composing a 
data science pipeline are executed by services that implement big data operators. The objective is to execute as just in time edge-based processes (similar to lambda functions), and they interact with the VDC underlying services only when the process to execute needs more resources.  This means that services are running on the edge, on processing entities with different computing and storage capacities. They can totally or partially execute their tasks on the edge and/or on the VDC. 
This in turn, creates the need for novel resource management approaches in streaming-based data science pipelines. These approaches should support and satisfy the data management strategy and stream exchange model between producers and consumers, invoke tasks with the underlying exchange model constraints on the compute and storage resources in the suitable form and modality and meet multi-objective competing performance goals.
Next we describe the architecture of big data operators and we show how they interact with the VDC. 
Later we will introduce our resource management approach for JITA-4DS.

\vspace{-3mm}
\paragraph{Big Data/Stream producing and processing services}

\begin{figure}[t] 
\centering
\includegraphics[width=0.95\textwidth]{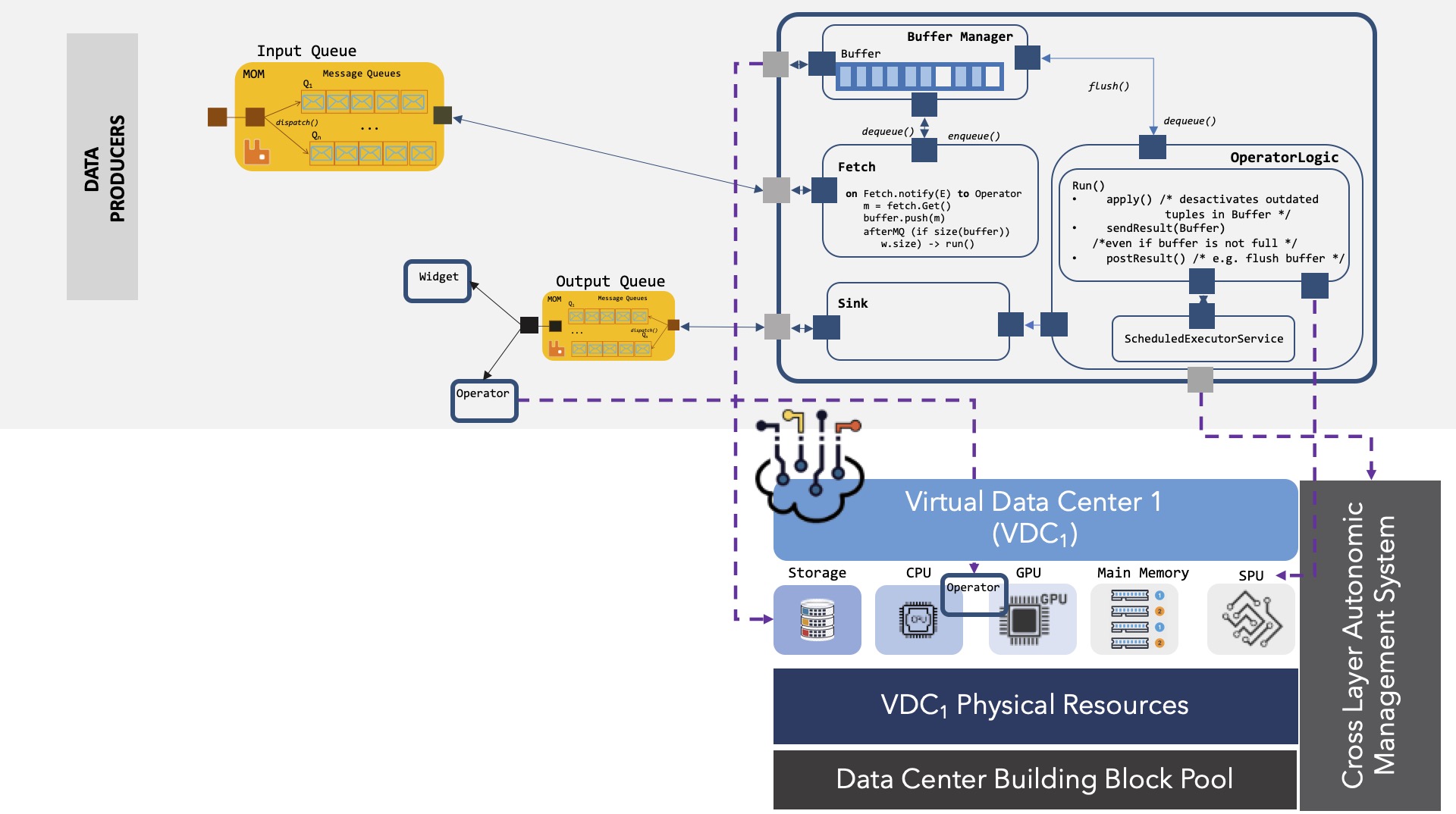}
\vspace{-2mm}
\caption{Architecture of a  Big data/stream processing service}
\label{fig:micro-service}
\vspace{-5mm}
\end{figure}

We assume that services that run on the edge produce and process data in batch or as streams. Data and stream processing services implement operators for supporting the analysis (machine learning, statistics, aggregation, AI) and visualization of big data/streams produced in IoT environments.
As shown in Figure \ref{fig:jita}, data and stream producing services residing on edge rely on underlying message-based communication layers for transmitting them to processing and storage services. These services can reside on edge or a VDC.
A data/stream service implements simple or complex analytics big data operations (e.g., fetch, sliding window, average, etc.). Figure \ref{fig:micro-service} shows the general architecture of a stream service.

The service logic is based on a scheduler that ensures the recurrence rate in which the analytics operation implemented by the service is executed. Stream/data processing is based on unlimited consumption of data ensured by the component {\sf\small Fetch} that works if streams are notiﬁed by a producer. This speciﬁcation is contained in the logic of the components {\sf\small OperatorLogic} and {\sf\small Fetch}.
%
As data is produced, the service fetches and copies the data to an internal buﬀer. Then, depending on its logic, it applies a processing algorithm and sends the data to the services connected to it. The general architecture of a service is specialized in concrete services implementing the most popular aggregation operations. 
These services can process data and streams on edge or on a VDC. 

Since RAM assigned to a service might be limited, and in consequence its buﬀer, every service implements a data management strategy by collaborating with the communication middleware and with the VDC storage services to exploit buﬀer space, avoiding losing data, and processing and generating results on time.
%
%
 Big stream/data operators combine stream processing and storage techniques tuned depending on the number of things producing streams, the pace at which they produce them, and the physical computing resources available for processing them on-line (on edge and VDC) and delivering them to consumers (other services).
  Stores are distributively installed on edge and on the VDC.
\vspace{-4mm}
\paragraph{Edge based Data Science (DS) Pipelines}
are expressed by a series of data processing operations applied to streams/data stemming from things, stores or services. A DS pipeline is implemented by mashing up services implementing operators based on a composition operation that connects them by expressing a data flow (IN/OUT data). Aggregation (min, max, mean) and analytics (k-means, linear regression, CNN) services can be composed with temporal windowing services (landmark, sliding) that receive input data from storage support or a continuous data producer, for instance, a thing.  The connectors are {\small\sf Fetch}, and {\small\sf Sink} services that determine the way services exchange data from/to things, storage systems, or other services (on-demand or continuous). Services can be hybrid (edge and VDC) services depending on the number of underlying services (computing, memory, storage) required.
To illustrate the use of a JITA-4DS, we introduce next a use case that makes full use of edge and VDC services configured ad-hoc for the analysis requirements.
\vspace{-4mm}
 \paragraph{Use Case: Analysing the connectivity of a connected society}
The experiment scenario aims at analyzing the connectivity of the connected society. 
The data set used was produced in the context of the  Neubot project\footnote{Neubot is a  project on measuring the Internet from the edges by the Nexa Center for Internet and Society at Politecnico di Torino (\url{https://www.neubot.org/})}.  It consists of network tests (e.g., download/upload speed over HTTP) realized by different users in different locations using and application that measures the network service quality delivered by different Internet connection types\footnote{The Neubot data collection was previously used in the context of the FP7 project S2EUNET.}.
%
%
 %
The type of queries implemented as data science pipelines  were the following:
\vspace{-2mm}
\begin{small}
\begin{verbatim}
EVERY 60 seconds compute the max value of download_speed 
of the last 3 minutes 
FROM 	cassandra database neubot series speedtests and streaming 
RabbitMQ queue neubotspeed

EVERY 	5 minutes compute the mean of the download_speed 
of the last 120 days 
FROM 	cassandra database neubot series speedtests and streaming 
rabbitmq queue neubotspeed
\end{verbatim}
\end{small}

For deploying our experiment, we built an IoT farm and implemented a distributed version of the IoT environment on a clustered version of RabbitMQ. This setting enabled to address a scaleup setting in terms of several data producers (things) deployed on edge. We installed aggregation operators as services distributed on the things and an edged execution environment deployed on servers deployed in different devices. The challenge is to consume streams and create a history of connectivity information and then combine these voluminous histories with new streams for answering the queries. 
Depending on the observation window size, the services access the observations stored as post-mortem data sets from stores at the VDC level and connect to on-line producers that are currently observing their connections (on edge). %
For example, the second query observes a window of 10 days size. 
Our services could deal with histories produced in windows of size 10 days or even 120 days. Such massive histories could be combined with recent streams and produce results at reasonable response times (order of seconds).

\vspace{-3mm}
\section{Preliminary experimental results}\label{sec:preliminary}

\subsection{Value of Service  based Scheduling and Resource Management}
JITA-4DS encourages a novel resource management methodology that is based on the time-dependent Value of Service (VoS) metric ~\cite{kumbhare2017value} to guide the assignment of resources to each VDC and achieve a balance between goals that usually compete with each other (e.g., completion time and energy consumption). VoS allows considering the relative importance of the competing goals, the submission time of the task (e.g., peak vs non-peak period), or the task's nature as a function of task completion time. A primary difference of our VoS metric from earlier studies on "utility functions" (e.g., \cite{khemka2015upe,briceno2011time,kargahi2011performance}) is the fact that we combine multiple competing objectives and we consider the temporal value of performing resource management at a given instant of time. This ability is crucial for meeting the SLO of edge-based data science pipeline execution, where the nature and amount of the data change dynamically among streams of data arriving from heterogeneous and numerous sets of edge devices.

\begin{wrapfigure}{r}{0.5\linewidth}
\centering
\includegraphics[width=0.9\linewidth]{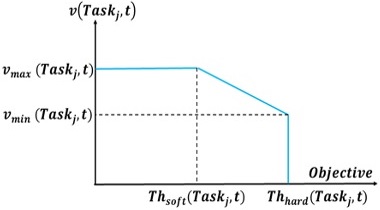}
\caption{General formulation for value vs. objective and thresholds.}
\label{fig:jita-stats}
\vspace{-4mm}
\end{wrapfigure}

In our earlier work~\cite{machovec2017utility}, we defined the value for a task as a monotonically-decreasing function of an objective (i.e., completion time, energy), illustrated in Figure \ref{fig:jita-stats}. The soft threshold parameter specifies the limit on an objective ($Th_{soft}$) until which the value earned by a task is maximum ($v_{max}$). Beyond the soft threshold, the value starts decreasing until the objective reaches the hard threshold. The hard threshold ($Th_{hard}$) specifies a limit on a given objective ($v_{min}$), beyond which zero value is earned. The linear rate of decay between $v_{max}$ and $v_{min}$ can be replaced by other functions if shown to provide an accurate representation.  

In Equation \ref{eq:one}, Task value ($V(Task_{j},t$)) represents the total value earned by completing task $j$ during a given period $t$. It is the weighted sum of earned performance and energy values based on Figure \ref{fig:jita-stats}. The $w_p$  and $w_e$  coefficients are used for adjusting the weight given to the performance and energy values. The importance factor $\gamma^{(Task_j)}$ expresses the relative importance among tasks. If either the performance function or energy function is 0, then the VoS is 0.

\begin{equation}\label{eq:one}
    V(Task_j,t)= (\gamma^{(Task_j )}) (w_p*v_p (Task_j,t)+w_e*v_e (Task_j,t))      
\end{equation}
                     
The VoS function, defined by Equation \ref{eq:two}, is the total value gained by n tasks in the workload that are completed during a given time period $t$.
\vspace{-3mm}
\begin{equation}\label{eq:two}
  VoS (t)=\sum_{j=1}^{n} V(Task_j,t)                                     
\end{equation}

The design of resource management heuristics for the JITA-4DS is a challenging problem. Resources must be interconnected and assigned to VDCs in a way that it will maximize the overall system VoS as defined in Equation \ref{eq:two}. 
Towards this goal, we designed a simulation environment and evaluated various heuristics by  experiments for a homogeneous environment much simpler than the JITA-4DS design illustrated in Figure~\ref{fig:jita}.
In the simplified environment, we studied the allocation of only homogeneous cores and memories from a fixed set of available resources to VMs, where each VM was for a single dynamically arriving task. Each task was associated with a task type, which has estimated execution time and energy consumption characteristics (through historical information or experiments) for a given number of assigned cores and assigned memory. To predict each application type's execution time and energy consumption, we use statistical and data mining techniques~\cite{kumbhare2017value,kumbhare2020dynamic,kumbhare2020value}, which represent the execution time and energy consumption as a function of the VDC resources. As an example, one of the heuristics was Maximum Value-per-Total Resources (Maximum VPTR). Its objective function is "task value earned / total amount of resources allocated," where the total amount of resources (TaR) for a task depends on the task execution time duration (TeD), the percentage of the total number of system cores (Cores) used and the percentage of the total system RAM used:

\begin{equation}
    TaR  = TeD \times  [(\% Cores ) +  (\%  RAM)]      
\end{equation}

\begin{figure}[t]
   \centering
   \includegraphics[width=\linewidth]{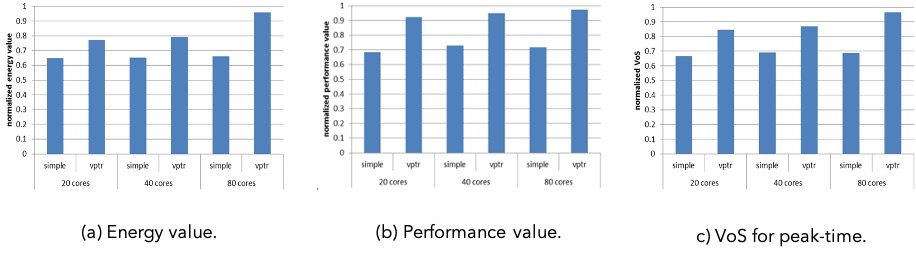}
   \vspace{-4mm}
 \caption{Value gain comparison of the VPTR over the simple heuristic}
 
   \label{fig:jita-stats-2}
   \vspace{-6mm}
 \end{figure}

We compare the VPTR (Value Per Total Resources) algorithm with a simple scheduling algorithm when applied to a workload that starts during peak usage time. For 80 cores, VPTR is able to have an improvement of almost 50\% in energy value and 40\% in performance value as shown in Figures \ref{fig:jita-stats-2} (a) and \ref{fig:jita-stats-2} (b), respectively. 
Figure \ref{fig:jita-stats-2} (c) shows the VoS when we combine both performance and energy values. Because the workload involves a peak system period, the Simple heuristic cannot efficiently utilise the resources, resulting in VPTR having up to 71\% increase in normalized VoS.

In general, each of these percentages can be weighted
for a given system, depending on factors such as the relative cost of cores and memory or relative demand for cores versus memory among applications in typical workloads. Each time the heuristic is invoked, it considers tasks that have arrived in the system but have not been assigned resources. For each task, independently, for each of its allowable resource configurations that will provide non-zero task value, we find the cores/memory that maximizes the task’s VPTR. Then, among all mappable tasks, we select the task with the maximum VPTR, call this task \emph{m}, and make that assignment of resources to task \emph{m}, or if the allocation is for a future time, we create a place-holder for task \emph{m}. We then update the system state information based on this assignment or place-holder and remove task \emph{m} from the list of mappable tasks. 

\subsection{JITA-4DS Simulation}

Because a composable system is a tightly integrated system, it will create new challenges in modeling application performance and designing resource allocation policies to maximize productivity. This creates the necessity for creating a simulation environment to experiment and evaluate composable systems and use them to design new resource allocation policies. 
We validated our simulation framework's functionality against emulation-based experiments based on a system composed of 64 nodes. Each compute node is a dual-socket with 125 GB of memory and InfiniBand QDR for the network interface. Each socket on the node has a CPU (Ivy Bridge-EP) with twelve cores and a maximum frequency of 2.40 GHz. The TDP of each CPU is 115 watts in our system. We access the power-specific registers on each CPU to monitor and control the CPU power consumption. On this system, we collect empirical power-performance profiles for our test applications and create models.  
We use the publicly available benchmarks from the NAS parallel benchmark suite (NPB)~\cite{NAS_NPB}. In our study, a synthetic workload trace is a list of jobs in the order of arrival time. We use the benchmarks listed in Table \ref{tab:nas_bm} to form a job in the workload trace. 
Each job entry in the workload trace consists of job arrival time, job name, maximum job-value, job's input problem size, iteration count, node configuration range, soft threshold, and hard threshold. 
We experimentally select the sampling range for these parameters to ensure our unconstrained HPC system is oversubscribed. We create offline models for each hybrid benchmark by using the modeling technique discussed in \cite{kumbhare2020dynamic}. 
\begin{table}[t]
 \fontsize{10pt}{12pt}\selectfont
 \fontsize{6pt}{7pt}\selectfont
\caption{NAS benchmarks used in this study.}
\vspace{-1mm}
\label{tab:nas_bm}
\centering
  \begin{tabular}{|c|c|c|c|}
    \hline
     \textbf{Benchmark} & \textbf{Description} & \textbf{MPI} & \textbf{MPI+OpenMP}\\\hline
    $CG$ & conjugate gradient &  \cmark & \xmark \\\hline
    $EP$ & embarrassingly parallel &  \cmark & \xmark \\\hline
    $FT$ & Fourier transform &  \cmark & \xmark \\\hline
    $IS$ & integer Sort &  \cmark & \xmark \\\hline
    $MG$ & multi-grid &  \cmark & \xmark \\\hline
    $LU$ & lower-upper Gauss-Seidel solver &  \cmark & \cmark \\\hline
    $BT$ & block tri-diagonal solver &  \cmark & \cmark \\\hline
    $SP$ & scalar penta-diagonal solver &  \cmark & \cmark \\\hline
\end{tabular}
\vspace{-4mm}
\end{table}

\begin{figure}[t]
\centering
  \mbox{
  \subfigure[$system\mbox{-}value\ from\ simulation$]{\label{fig:emsim_value_comp_0} \includegraphics[scale=.35]{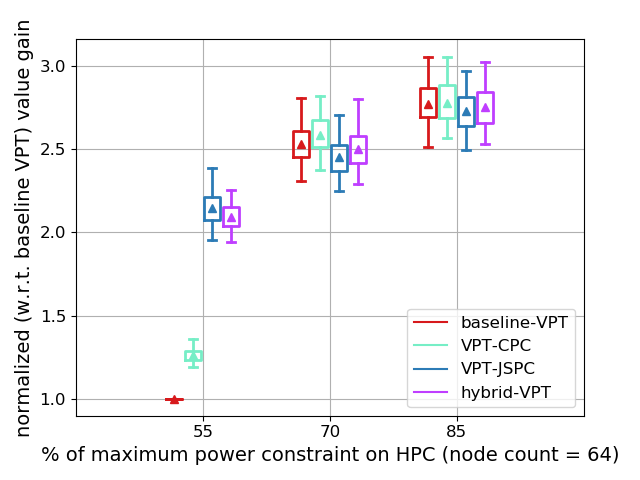}}
  \subfigure[$system\mbox{-}value\ from\ emulation$]{\label{fig:emsim_value_comp_1} \includegraphics[scale=.35]{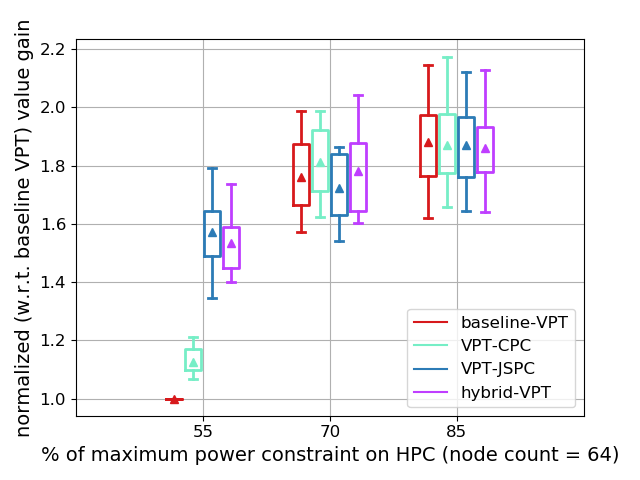}}
   }
   
  
  \vspace{-3mm}
  \caption{Comparing the simulation and emulation results.}
 \label{fig:emsim_comp}
 \vspace{-4mm}
\end{figure}

Similar to our emulation HPC prototype, we simulate a system composed of 64 nodes. We use a similar set of hybrid benchmarks and their models to create workload traces. For the simulation study, we create 50 workload traces, and each trace is composed of 1000 jobs in the order of their arrival time. Each trace simulates a workload of approximately 50 hours at three different system-wide power constraints (55\%, 70\%, and 85\%). While simulating a workload trace under a given power constraint, we assume our model prediction error is zero for each benchmark.  In Figures \ref{fig:emsim_value_comp_0} and \ref{fig:emsim_value_comp_1}, we compare the system-value earning distribution from our emulation and simulation studies respectively using a set of value-based heuristics covering  Value-Per-Time (baseline-VPT) and its variations~\cite{kumbhare2020dynamic,kumbhare2020value} Common Power Capping (VPT-CPC), Job-Specific Power Capping (VPT-JSPC), and hybrid that combines  VPT-CPC and VPT-JSPC. Even though our power-aware algorithms' normalised system-value earnings is higher in the simulations compared to the emulations, we observe a similarity in the pattern of the system-value earnings of the algorithms as the power constraint is relaxed from 55\% to 85\%. The reason for differences in the magnitudes of system-value earnings can be attributed to our simulation study, assuming all the system's CPUs are identical.

\vspace{-3mm}
\paragraph{Discussion}
In the JITA-4DS environment, the resource management problem is much more complex and requires the design of new heuristics. The computing resources allocated to the VDC for a given class of applications are a heterogeneous mixture of different types of processing devices (different CPUs, different GPUs, different accelerators, etc.) with various execution performance and energy consumption characteristics. They depend on each of the specific applications being executed by that VDC. 
For example, several aspects remain open, like the ad-hoc design of the JITA-4DS resource management system for a  VDC built from a fixed set of components. The design of a JITA-4DS instance is determined by the execution time and energy consumption cost, and resources requirements of a data science pipeline. Therefore, it is necessary to identify the configuration choices for a given pipeline and define VDC resources' effective resource allocation strategies dynamically.   In general, for determining the dynamic resources requirements of data science pipelines at runtime, it is necessary to consider two challenges.
First, calculate a VDC-wide VoS for a given interval of time,  weigh individual values of various instances of pipelines.  
Second, propose objective functions that can guide heuristics to operate in the large search space of resource configurations. The objective is to derive possible adequate allocations of the shared and fixed set of VDC resources for several instances of data science pipelines.
    
    
 %
 We have observed that regarding the resource management system for JITA-4DS,  decisions must be made to divide the shared, fixed resource pool across different VDCs to maximize the overall system-wide VoS. 
 All of the above single VDC challenges still apply and interact across VDCs. Additional problems, such as determining when resources should be reallocated across VDCs and do so in an on-line fashion, must be addressed. This includes the methodologies for reassigning resources that do not interfere with currently executing applications on different VDCs affected by the changes and measuring and accounting for the overhead of the establishment of new VDC configurations. 
 
\vspace{-3mm}
\section{Conclusion and Future Work}\label{sec:conclusion}
\vspace{-3mm}
This paper introduced JITA-4DS, a virtualised architecture that provides a disaggregated data center solution ad-hoc for executing DS pipelines requiring elastic access to resources. DS pipelines process big streams and data coordinating operators implemented by services deployed on edge. Given that operators can implement greedy tasks with computing and storage requirement beyond those residing on edge, they interact with VDC services. We have set the first simulation setting to study resources delivery in JITA-4DS.


We are currently addressing challenges of VDCs management on simpler environments, on cloud resource management heuristics, 
big data analysis, 
and data mining for performance prediction.
To simulate, evaluate, analyze, and compare different heuristics we will build 
simulators for simpler environments and combine open source simulators for different levels of the JITA-4DS hierarchy.

%
%
%
 \bibliographystyle{splncs04}
 \bibliography{biblio}

\end{document}